\documentclass[conference]{IEEEtran}

\usepackage[left=0.673in,right=0.673in,top=0.755in,bottom=1.050in]{geometry}
\usepackage{graphicx}
\usepackage{cite}
\usepackage{amsmath,amssymb,amsfonts}
\usepackage{algorithmic}
\usepackage{graphicx}
\usepackage{textcomp}
\usepackage{xcolor}
\usepackage{multirow}
\usepackage{subfig}
\usepackage[inkscapelatex=false]{svg}

\usepackage{balance}

\usepackage{url}

\usepackage{listings}
\lstalias{plantuml}{}
\usepackage{xcolor}

 \definecolor{codegreen}{rgb}{0,0.6,0}
\definecolor{codegray}{rgb}{0.5,0.5,0.5}
\definecolor{codepurple}{rgb}{0.58,0,0.82}
\definecolor{backcolour}{rgb}{0.95,0.95,0.92}
 
\lstdefinestyle{mystyle}{
    backgroundcolor=\color{backcolour},   
    commentstyle=\color{codegreen},
    keywordstyle=\color{magenta},
    numberstyle=\tiny\color{codegray},
    stringstyle=\color{codepurple},
    basicstyle=\ttfamily\footnotesize,
    breakatwhitespace=false,         
    breaklines=true,                 
    captionpos=b,                    
    keepspaces=true,                 
    numbers=none,                    
    numbersep=5pt,                  
    showspaces=false,                
    showstringspaces=false,
    showtabs=false,                  
    tabsize=2
}
 
\def\BibTeX{{\rm B\kern-.05em{\sc i\kern-.025em b}\kern-.08em
    T\kern-.1667em\lower.7ex\hbox{E}\kern-.125emX}}

\begin{document}

\title{A Cyber-Twin Based Honeypot for Gathering Threat Intelligence}


 \author{

     \IEEEauthorblockN{Muhammad Azmi Umer\IEEEauthorrefmark{1}, Zhan Xuna \IEEEauthorrefmark{1}, Yan Lin Aung \IEEEauthorrefmark{2}, Aditya P. Mathur \IEEEauthorrefmark{1}, Jianying Zhou 
     \IEEEauthorrefmark{1}}

     \IEEEauthorblockA{\IEEEauthorrefmark{1}Singapore University of Technology and Design, Singapore
     \\\{azmi\_umer, aditya\_mathur, jianying\_zhou\}@sutd.edu.sg, zhan\_xuna@mymail.sutd.edu.sg}
     \IEEEauthorblockA{\IEEEauthorrefmark{2}University of Derby, United Kingdom
     \\y.aung@derby.ac.uk}
     }


\maketitle


\begin{abstract}
Critical Infrastructure (CI) is prone to cyberattacks. Several techniques have been developed to protect CI against such attacks. In this work, we describe a honeypot based on a cyber twin for a water treatment plant. The honeypot is intended to serve as a realistic replica of a water treatment plant that attracts potential attackers. The attacks launched on the honeypot are recorded and analyzed for threat intelligence. The intelligence so obtained is shared with the management of water treatment plants, who in turn may use it to improve plant protection systems. The honeypot used here is operational and has been attacked on several occasions using, for example, a ransomware attack that is described in detail.
\end{abstract}


\section{Introduction}
\label{sec:intro}
The integration of Critical Infrastructure such as water treatment and distribution plants and the electric grid into the public network poses significant threats. Maroochy sewage spills\,\cite{slay2008lessons}, Stuxnet\,\cite{falliere2011w32}, and colonial pipeline attack\,\cite{colonial} are  examples of such attacks. Various studies are available in the literature to protect CI. Such studies are mainly intrusion detection systems (IDS), as evident from a survey\,\cite{umer2022machine}.  

\begin{figure*}
\centering
  \includegraphics[width=\textwidth, height=8cm]{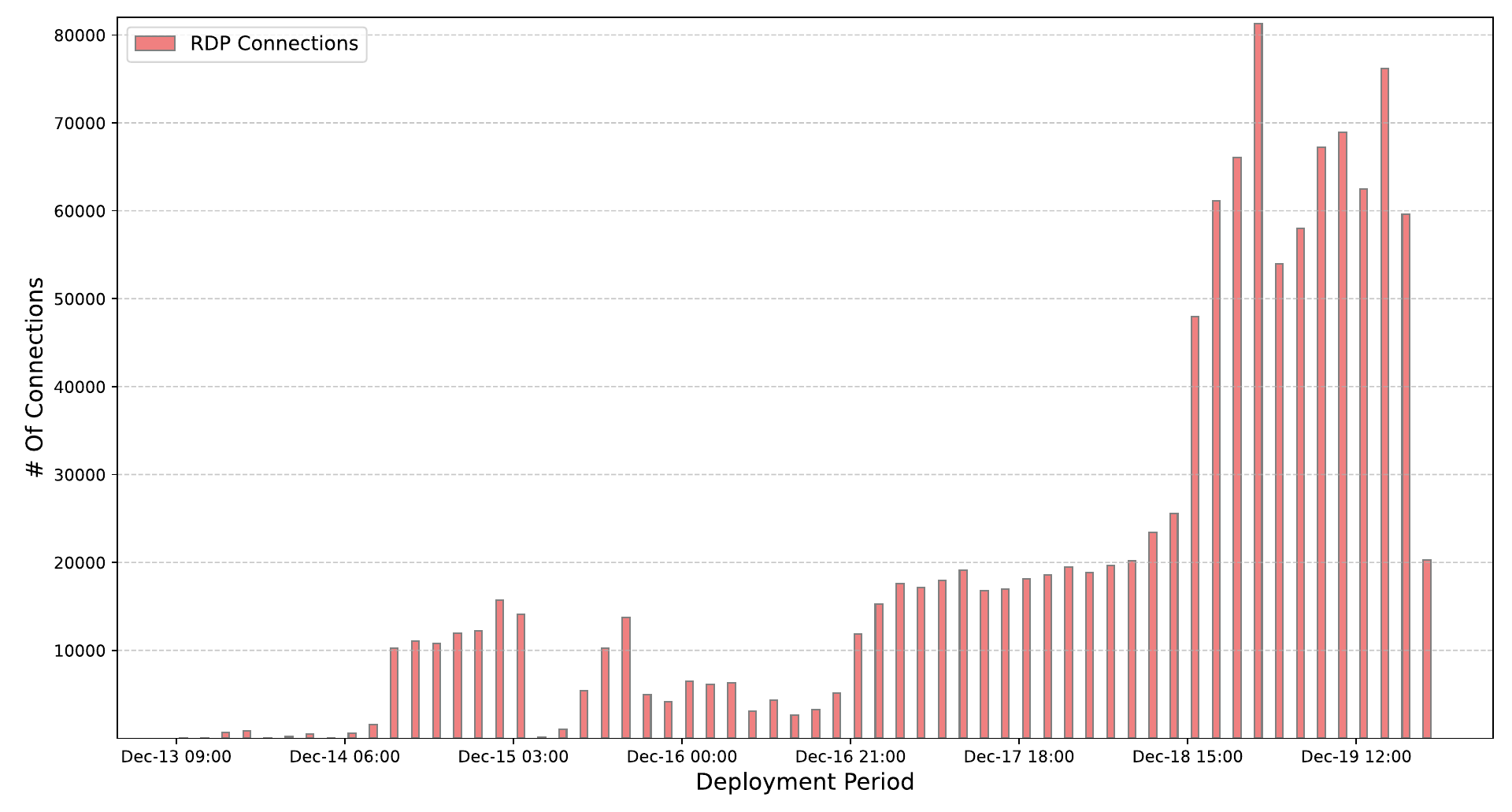}
   \caption{Number of RDP Connections during the deployment period}
  \label{fig:RDP}
\end{figure*}

\begin{figure}
  \centering 
  \includegraphics[width=10cm, height=5cm]{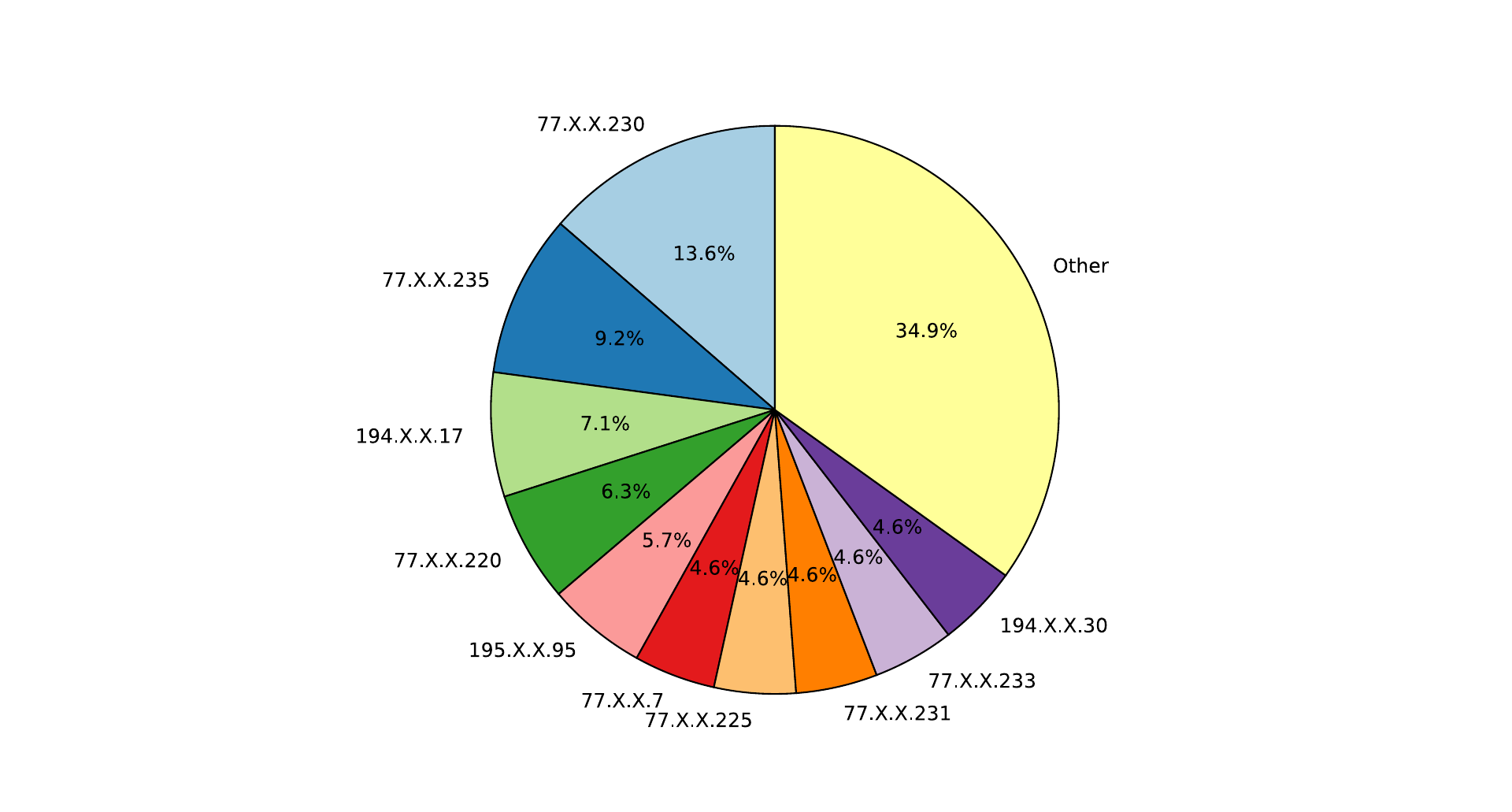}
   \caption{IP addresses which connected to the honeypot during the deployment period}
  \label{fig:IncomingIPs}
\end{figure}

The approach proposed here is to assess threats to the CI while isolating it from the attacker. Toward this end, we have deployed a cyber twin of the Water Treatment Plant (WTP) as the honeypot. There are two main components of the honeypot. One is the twin of the WTP  and second is the Human Machine Interface (HMI) which communicates with the twin. We have deliberately allowed the access of HMI via Remote Desktop Protocol (RDP) through a weak password. Once an attacker accesses the system, it appears as a real HMI of WTP. However, an attacker cannot access the twin due to secure network configuration. Figure~\ref{fig:RDP} shows the number of connections made through RDP, while Figure~\ref{fig:IncomingIPs} shows the incoming IPs to the honeypot during the deployment period. This deployment remained live for 9-days. During this period, 11 successful logins (breaches) were recorded in the system. Among these, one of the successful logs indicated the launch of the ransomware attack on the twin. This paper discusses the ransomware attack, communication with the  attacker, and how the proposed topology was successful in making the twin safe and secure despite successful breaches.\newline 

\noindent {\em Digital and cyber-twins}: The term ``digital twin" often refers to  a near software replica of a physical plant. A cyber twin is also a near software replica of a physical plant but with cyber-security as the focus. Thus, while a digital twin focuses on tasks such as operator training and process evaluation, a cyber twin focuses on evaluating the impact of cyber attacks on the twinned plant and training to defend the plant when under cyber-attack. A digital twin often uses virtual reality  to show the plant twinned for training whereas a cyber twin uses Human Machine Interface (HMI) to show the plant state.
\vskip0.1in
\noindent {\em Novelty}: There exist several honeypots intended to obtain threat intelligence. The work described here is novel in that it uses a cyber, not a digital,  twin of an operational water treatment plant. The plant itself is not disclosed in this work, though its architecture is described briefly. While references to  ICS-based honeypots are found in the literature, the work presented here is perhaps the first fully functional cyber-twin deployed as a honeypot.\newline 

\noindent {\em Contributions}: The key contributions of our work are enumerated below.

\begin{itemize}
    \item The deployment of a cyber twin-based honeypot that represents  an industrial replica of a water treatment plant in a  public environment.

    \item An architecture for the secure deployment of critical infrastructure.

    \item Analysis of a ransomware attack.

    \item Communication with the ransomware attacker to understand the potential risks and the associated benefits.\newline
\end{itemize}

\noindent {\em Organization}: The remainder of this work is organized as follows. Section\,\ref{sec:rw} summarizes literature focused on honeypots based on cyber twins. The water treatment plant on which our cyber twin is based is briefly described in Section\,\ref{sec:SWaT}; details and citations are omitted on purpose. Honeypot deployment, including its components and the pathway to access the twin is discussed in Section\,\ref{sec:honeypot_deployment}. The ransomware attack on the honeypot, followed by communication with the attacker, is discussed in Section\,\ref{sec:ransomware_attack}. Section\,\ref{sec:discussion} summarizes the lessons learned and recommendations based on our work. Section\,\ref{sec:conclusion} contains the conclusions from our study.

\begin{table*}
\centering
\begin{tabular}{c c c c c}
 \hline
 \textbf{References} & \textbf{Domain} & \textbf{Real-World Deployment} & \textbf{Real-Attacks} & \textbf{Received Ransomware Attack}\\
\hline
\hline
 Proposed Work &  Water Plants & $\checkmark$ & $\checkmark$ & $\checkmark$ \\
 
 \cite{nintsiou2023threat} & Not specified & $\times$ & $\times$ & $\times$ \\

  \cite{liatifis2024sihoneypot} & Autonomous Vehicles & $\times$ & $\times$ & $\times$ \\

\cite{yigit2023twinpot}  & Maritime & $\times$ & $\times$ & $\times$ \\

\cite{krishnaveni2024cyberdefender} & Industrial CPS  & $\checkmark$ & $\checkmark$ & $\times$ \\
 \hline
 \hline
\end{tabular}
\caption{Comparison with cyber twin based honeypots}
 \label{tab:related_work} 
\end{table*}

\section{Related Work}
\label{sec:rw}
A framework for the honeypot based on a digital twin was proposed in~\cite{nintsiou2023threat}. This framework provides current and past network information to update the real-honeypot and assist human operators. However, there does not exist an implementation of the proposed framework. Similarly, a digital twin-based honeypot for autonomous vehicles was proposed in~\cite{liatifis2024sihoneypot}. Its architecture was evaluated using synthetic attacks. However, there is no implementation of the proposed architecture references in the literature. In contrast, our proposed architecture was deployed in a public environment and includes an analysis of real-world attacks, including a ransomware attack. In~\cite{krishnaveni2024cyberdefender} a CyberDefender framework is proposed to analyze attacks on digital twin-based Industrial Cyber Physical Systems (ICPS).  The authors demonstrated a Proof of Concept (PoC) for the framework by deploying the T-Pot honeynet on AWS via Mininet-enabled MiniCPS. Another study\,\cite{zimba2018multi} modeled the multi-crypto ransomware attacks. A cascaded network segmentation approach using DMZ was proposed to prioritize the security of production network devices followed by the SCADA networks. While the study proposed here is protecting the twin of a an operational plant by exposing the Human Machine Interface (HMI) of the plant to the attacker. Doing so compromises  the security of the HMI while protecting the twin.   

An ICS honeypot based on Conpot was developed in\,\cite{pliatsios2019novel}. This honeypot mimics the Modbus-based physical device by using network traffic captured from real devices. To generate network traffic, a custom HMI emulator was developed that interacts with the honeypot. Similarly, a honeypot based on the water treatment testbed was proposed in\,\cite{antonioli2016towards}. It is an  interactive Ethernet/IP based ICS honeypot developed without the use of complete virtualization technologies. The study presented here uses an operational Water Treatment Plant (WTP) as a cyber twin to setup the honeypot. The study conducted in\,\cite{navarro2019gathering} presents the development of a honeypot based on a water treatment plant. That study was conducted from the perspective of a cyber security service provider who has been tasked to secure  critical infrastructure. The honeypot infrastructure consists of the ICS, simulation system, and the infrastructure for  monitoring of cyber activity. An integrated framework for the design, implementation, and evaluation of adaptive honeypots within Internet of Things (IoT) environments was proposed in\,\cite{morozov2024sweet}. The architecture proposed in this work adjusts its behavior based on the observed attack patterns. A high-interaction Windows-based honeypot for the enterprise environment was proposed in~\cite{aung2025honeywin}. They recorded 354 successful logins via RDP during the deployment period of 34 days. The proposed honeypot was used to initiate a Simple Mail Transfer Protocol (SMTP) brute-force bot attack through an SSH session.

\begin{figure*}
\centering
  \includegraphics[width=0.7\textwidth, height=7cm]{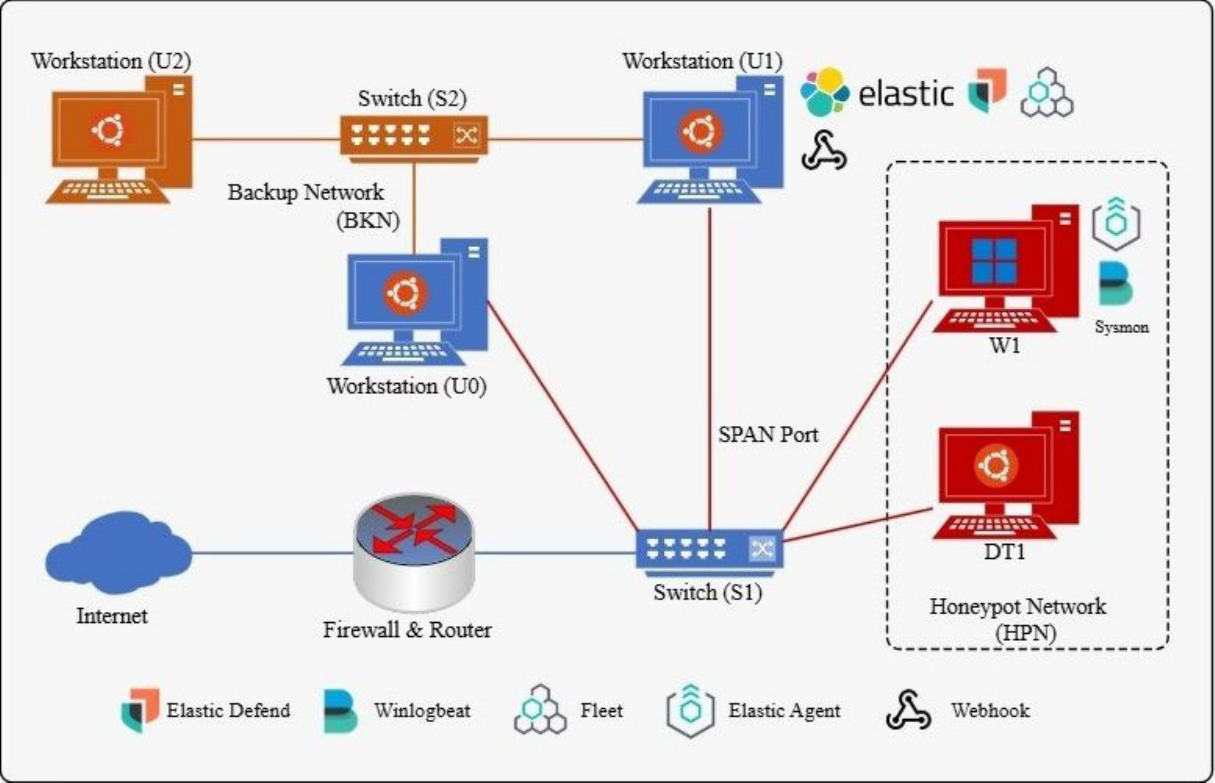}
  \caption{Components of the honeypot that includes the cyber twin of a water treatment plant.}
  \label{fig:honeywin_dt}
\end{figure*}

\section{The Twinned Water Treatment Plant}
\label{sec:SWaT}
In this study, we utilized the cyber twin of an operational  Water Treatment Plant. This cyber twin was developed by one of the authors to support education and research. It has been in use since 2021. In the remainder of this paper the plant twinned is referred to as WTP. The WTP is a six-stage plant capable of producing 5\,gallons of treated water per minute during 24/7 autonomous operation. Each stage is managed by a Programmable Logic Controller (PLC) and features sensors for measurement (e.g., LIT101 for tank T101 level) and actuators for control (e.g., MV101 for flow into T101). The WTP employs 68 sensors and actuators. Some actuators, such as pump P102, are  backups  activated only when the primary actuator (e.g., P101) fails.

The plant control room contains a SCADA workstation that provides comprehensive data and control access to plant components. Plant engineers can monitor and adjust process parameters through this workstation and a Human Machine Interface (HMI). Control code for each PLC can be loaded via the workstation. A historian records process states and network traffic at predefined intervals. The WTP's multi-layer network facilitates communication between all components. Level 0 utilizes a ring topology for PLC-sensor/actuator communication within each stage. Level 1 utilizes a star topology for communication between PLCs, SCADA, the HMI, and the historian. Both wired and wireless connections are supported at Level 1 and for sensor communication at Level 0. Plant operation is initiated and controlled via the SCADA workstation. State information for each sensor and actuator is displayed on the workstation and HMI, and also logged in the historian. Anomaly detectors developed by the researchers record alerts and provide visual alerts and messages to operators. The WTP is vulnerable to network attacks at all levels and to direct access to PLCs, SCADA, and the HMI. Physical attacks, such as sensor tampering or wire disconnection, are also possible.

\begin{figure*}
\centering
  \includegraphics[width=0.9\textwidth]{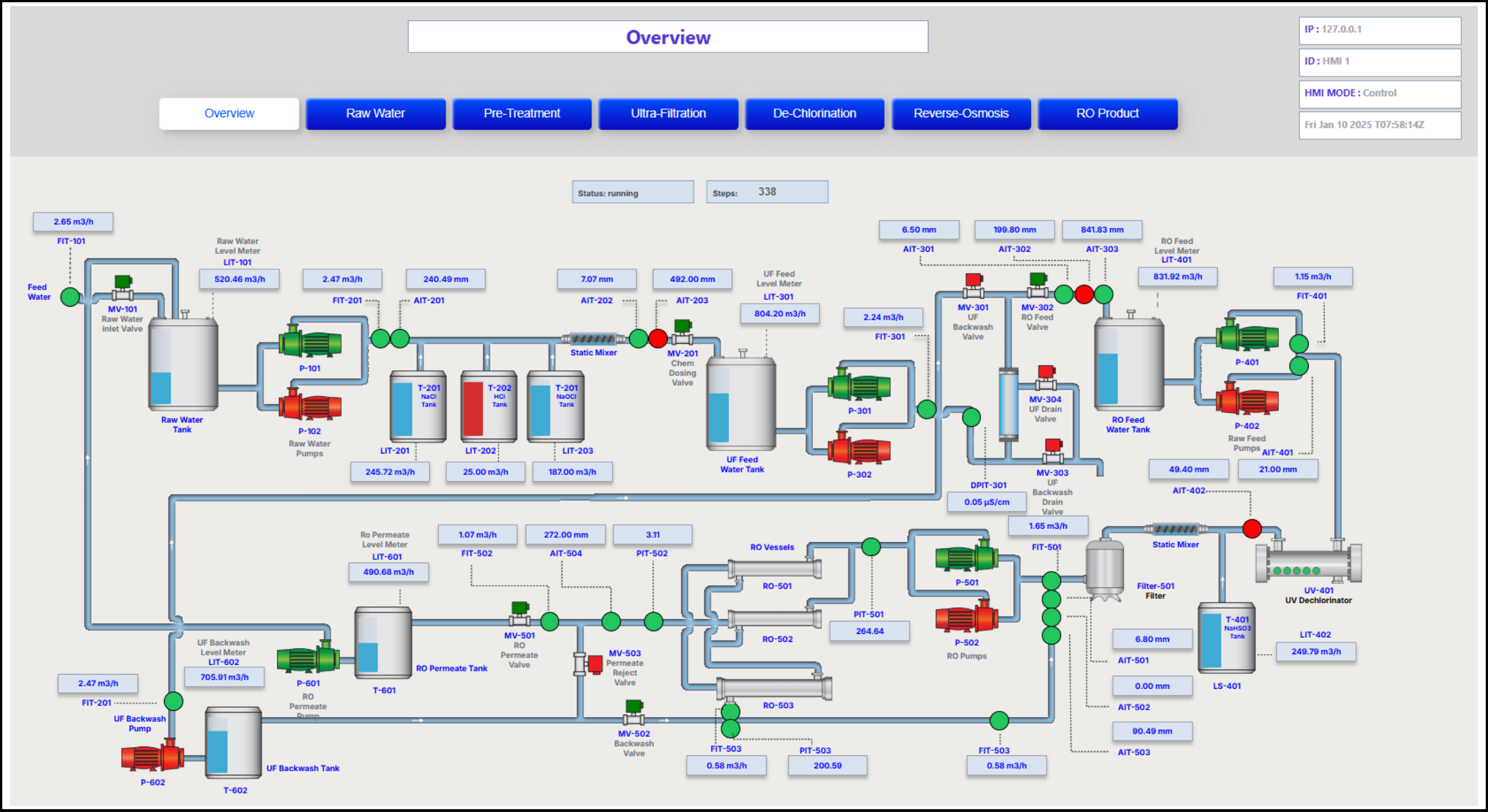}
  \caption[Caption for LOF]{HMI  of the  cyber twin.}
  \label{fig:hmi}
\end{figure*}

\section{Honeypot Deployment}
\label{sec:honeypot_deployment}
The honeypot used in our work consists of several interconnected components. The overall architecture of the honeypot is shown in Figure~\,\ref{fig:honeywin_dt}. The cyber twin of WTP was deployed as a honeypot system. The threat model developed in such a way that an employee of an organization is remotely accessing a work personal computer (PC) via a VPN connection. To realize such a scenario, we decided to use a PC with Windows 11 Professional rather than the Windows Server. Each component in the honeypot is described next. 

\subsection{Components in the Honeypot}

 \noindent {\em Workstation (W1)}: This workstation runs Windows 11 and has internet access. Incoming and outgoing network traffic to and from it (W1) are captured separately via the SPAN port. Windows host events on W1 are logged using Sysmon. The W1 host provides a Human Machine Interface (HMI) as the web application frontend for the WTP.\newline

\noindent{\em Workstation (U0)}: This workstation captures the incoming network traffic passing through the firewall and router.\newline

\noindent {\em Workstation (U1)}: Outgoing network traffic from Switch (S1) is captured using the SPAN port and stored as PCAP files on workstation (U1). An ELK stack is set up on U1 for log management and analysis. An Elastic Agent (EA) is installed on W1 to collect and transfer Syslogs to U1.\newline

\noindent {\em Workstation (U2)}: This workstation acts as the backup server for the incoming traffic captured on U0. Furthermore, it stores the outgoing network traffic originating from S1 and Windows log events from U1. It possesses the capability to access U0 and U1, while access in the reverse direction is restricted.\newline

\noindent {\em Workstation (DT1)}: This workstation contains the cyber twin of WTP described briefly n Section~\ref{sec:SWaT}. DT1 does not have the Internet access as controlled by the firewall rule inside the Firewall and Router. DT1 only accepts connections from W1 via a specific IP address and protocol, a websocket connection in this case. Workstation U2 has Secure Shell (SSH) access with authorized public key via a dedicated Backup Network (BKN) link to backup the logs from DT1. This integration enables to monitor network traffic and host logs of W1, and provides real-time email alerts upon a successful login (i.e, breach). The security of DT1 is hardened and access is tightly controlled.\newline

\noindent {\em Switch (S1)}: The SPAN port on this switch is used to capture outgoing network traffic and store it as PCAP files on U1.\newline

\noindent {\em Switch (S2)}: This switch is installed on the backup network to provide the backup server (U2) access to incoming network traffic captured at U0. It also provides U2 access to outgoing network traffic and Windows event logs captured on U1.\newline

\noindent {\em Firewall and router}: This firewall and router are used to configure DNS settings, define firewall rules, perform DHCP reservations, and manage VLAN trunks. Outgoing connections from the Honeypot Network (HPN) are blocked by default. The firewall rules allow internet access only to selected devices.\newline

\noindent {\em Backup network (BKN)}: The backup network provides the backup server (U2) with connectivity to U0 and U1. It does not have an internet connection.\newline 

\noindent {\em ELK Stack}: The ELK Stack was integrated into U1. It automates Elastic search indices and performs log management and analysis.\newline




\begin{figure}
\centering
  \fbox{\includegraphics[width=8cm, height=9cm]{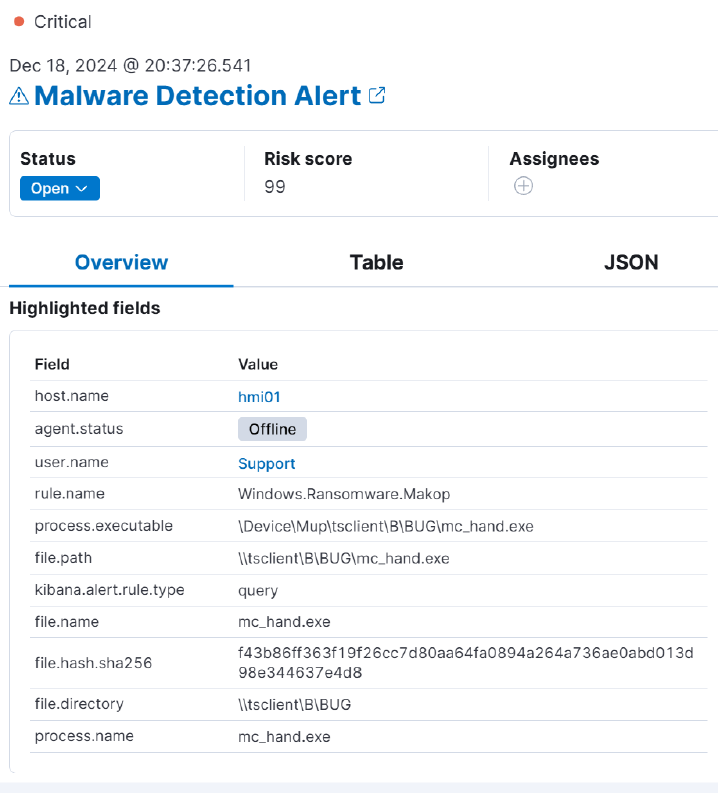}}
  \caption{Attributes of the malware launched on the honeypot. }
  \label{fig:malware}
\end{figure}

\subsection{Pathway to Access the Twin}
\label{subsec:pathway}

When the attackers successfully log in through RDP, they gains access to the workstation, W1, as shown in Figure~\ref{fig:honeywin_dt}.
They are able to launch the HMI through a shortcut placed on the desktop.
The HMI, as depicted in Figure~\ref{fig:hmi}, functions similar to a real-world HMI.
Using the HMI attackers send commands to the cyber twin, which is installed on a separate workstation. 
The twin receives these commands and executes the corresponding actions.
The HMI continuously receives updated plant state from the twin.

Importantly, access to the workstation hosting the twin is tightly controlled via a host firewall.
It allows sending and receiving data on a specific protocol.
This design and code-level security measures help protect the twin from direct attacker access.



\section{Ransomware Attack on Honeypot}
\label{sec:ransomware_attack}
Ransomware attacks typically follow a certain pattern and can be categorized into multiple stages based on the MITRE ATT\&CK framework, including Initial Access, Execution, Persistence, Defense Evasion, Discovery, Lateral Movement, and Command and Control. In the Initial Access stage, attackers often gain entry to a target system through phishing emails or brute-force attacks. Our honeypot system attracts attackers by exposing an RDP port to the public and creating user accounts with relatively weak credentials. Once an attacker gets access to the system, then attackers attempt to expand their attack surface, often leveraging publicly available open-source tools for lateral movement and system discovery. In our case study, the attacker disabled Windows Defender notifications to evade detection, made multiple attempts to laterally move and access the actual HMI system, and employed malicious software to gather network and process information.
During the execution phase of ransomware deployment, attackers typically seek sensitive files before initiating encryption. Once a system is deemed a worthwhile target, ransomware is deployed, often creating multiple redundant encryption processes to prevent termination. In subsequent stages, attackers may use legitimate external web services for data exfiltration or command and control (C2), leveraging popular websites to evade detection.
Our HoneyTwin experiment captured a complete ransomware attack chain. Fortunately, the attacker only encrypted the SCADA system's web access link, avoiding direct compromise of the SCADA system itself. By simulating a realistic SCADA deployment, the honeypot successfully deceived the attacker and recorded all attack behaviors. These insights enable targeted improvements to real-world CPS security defenses, demonstrating the strategic value of deploying honeypots in CPS environments.

The ransomware attack was detected by Elasticsearch integrated with the honeypot, as shown in Figure~\ref{fig:malware}.
The system  was disconnected from the Internet to initiate forensic investigations.
The attacker encrypted multiple files and left a ransom note (or ``readme file'') on the host. 
The ransomware name was $"mc\_hand.exe"$ which is from $``Makop"$ ransomware family.

\subsection{Makop Ransomware}

Makop ransomware~\cite{makop} is an evolution of the PHOBOS strain and poses a significant threat to organizations, particularly those working in critical sectors, e.g., energy, water, or gas, etc. It encrypts sensitive data using robust AES-256 encryption, appends '.makop' or '.mkp' as an extension to the compromised files, and demands Bitcoin ransom for decryption~\cite{makop2}. This malware seeks infiltration of the network through vulnerabilities such as open RDP services, deceptive phishing campaigns, infected attachments, and compromised torrent downloads. To maximize damage, Makop employs tools such as PowerShell, Mimikatz, and PsExec for lateral movement and disables volume shadow copies which hinders recovery efforts. Operating under an affiliate model, Makop represents a serious and persistent cyber threat to critical infrastructures.


\subsection{Information Collection and Ransomware Execution by the Attacker}

\begin{figure}
  \includegraphics[width=9cm, height=5cm]{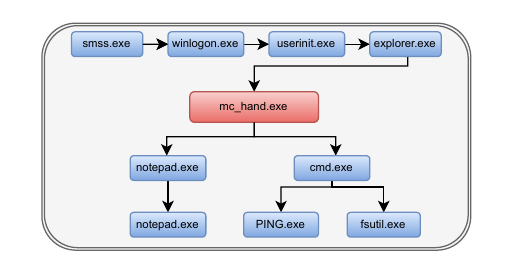}
  \caption{Process chain in the honeypot.}
  \label{fig:ProcessChain}
\end{figure}

The attacker created the executable file $``rew\_NS.exe"$. This file has ability to query information about shared network resources, reads the active computer name, queries process information and has ability to enumerate volumes.
The attacker also retrieved stored credentials from the Credential Manager. They may be attempting to obtain sensitive information such as usernames, passwords, etc. The attacker executed the $``mc\_hand.exe"$ file located in the \textbackslash \textbackslash tsclient\textbackslash B \textbackslash BUG\textbackslash directory. This path indicates that it is on a remote desktop shared path, and it is likely that the file was uploaded through a remote desktop session. Moreover, $``mc\_hand.exe"$ was executed dozens of times by $"explorer.exe"$ process. This is likely due to the persistence mechanism of ransomware, as ransomware often tries to execute itself multiple times to ensure it runs successfully, even if some instances fail. The process chain of a ransomware attack is shown in Figure\,\ref{fig:ProcessChain}. The operations of ransomware are divided into three parts: process operations, registry operations, and network connections.

\subsubsection{Process Operations}
The ransomware attacker created a process (Notepad.exe). It was was launched with the file +README-WARNING+.txt from the C:\textbackslash Users\textbackslash Support\textbackslash Desktop directory as shown in listing 1. The Notepad.exe was used to open a session file. This session can be used to track the status or communication of ransomware. The session could be related to the malware's C2 (Command and Control) server, and the ransomware indeed attempted to connect to another IP in the United States as shown in listing 2. The attacker also executed a series of operations as shown in listing 3. The ping 1.1.1.1 -n 5 command is likely used to test network connectivity and to ensure that their C2 (command and control) server is reachable. The “fsutil file setZeroData offset=0 length=131072 \textbackslash \textbackslash tsclient\textbackslash \textbackslash B\textbackslash \textbackslash BUG\textbackslash \textbackslash mc\_hand.exe” command can be used to clean the contents of the ransomware file. The del /q /f ``\textbackslash \textbackslash tsclient\textbackslash \textbackslash B\textbackslash \textbackslash BUG\textbackslash \textbackslash mc\_hand.exe" command deletes the ransomware file mc\_hand.exe, clearing attack traces and preventing it from being tracked or recovered.

\begin{lstlisting}[language=plantuml, caption=Process creation by ransomware]
``\"C:\\Program Files\\WindowsApps\\Microsoft.WindowsNotepad_11.2410.21.0_x64__8wekyb3d8bbwe\\Notepad\\Notepad.exe\`` \"C:\\Users\\Support\\Desktop\\+README-WARNING+.txt\"" \end{lstlisting}

\begin{lstlisting}[language=plantuml, caption=Session creation by ransomware]
``C:\ProgramFiles\WindowsApps\Microsoft.WindowsNotepad_11.2410.21.0_x64__8wekyb3d8bbwe\Notepad\Notepad.exe" /SESSION:igsf1t3JF0mJYAAZ9bmc2wEtQwA6AFwAVQBzAGUAcgBzAFw
    AUwB1
AHAAcABvAHIAdABcAEQAZQBzAGsAdABvAHAAXAArAFIARQBBAEQA
TQBFAC0AVwBBAFIATgBJAE4ARwArAC4AdAB4AHQAAAAAAAAAAAAA
AAAAAAAAAAAAAAAAAAAAAAAAAAAAAAAAAAAAAAAAAAEKAAGI
SQAA3ckXSQADAAAAAAAA. \end{lstlisting}

\begin{lstlisting}[language=plantuml, caption=Series of operations executed by the ransomware process]
``C:\\WINDOWS\\system32\\cmd.exe",
        "/c",
        "ping",
        "1.1.1.1",
        "-n",
        "5",
        "&",
        "fsutil",
        "file",
        "setZeroData",
        "offset=0",
        "length=131072",
        "\\\\tsclient\\B\\BUG\\mc_hand.exe",
        "&",
        "del",
        "/q",
        "/f",
        "\\\\tsclient\\B\\BUG\\mc_hand.exe" \end{lstlisting}

\subsubsection{Registry Operations}
The attacker also performed the registry operations. For example, the ransomware modified the registry key so that its traffic bypasses proxy settings, thus evading network monitoring as shown in listing 4. The ransomware also modified the settings to ensure the system automatically selects its designated proxy server as shown in listing 5. The ransomware also modified the path used to store system-level settings and configurations as shown in listing 6. It's modification can affect the behaviour of notifications, such as certain warnings or pop-up alerts can not be displayed. The attacker replaced the wallpaper by using their file 4506.tmp.bmp as shown in listing 7.

\begin{lstlisting}[language=plantuml, caption=Modification in the registry key by the ransomware to bypass proxy settings ]
"HKU\\S-1-5-21-4252838199-3690154089-1328794218-1003\\Software\\Microsoft\\Windows\\CurrentVersion\\Internet Settings\\ZoneMap\\ProxyBypass \end{lstlisting}

\begin{lstlisting}[language=plantuml, caption=Modification in the registry key by the ransomware to select designated proxy server ]
"HKU\\S-1-5-21-4252838199-3690154089-1328794218-1003\\Software\\Microsoft\\Windows\\CurrentVersion\\Internet Settings\\ZoneMap\\AutoDetect" \end{lstlisting}

\begin{lstlisting}[language=plantuml, caption=Modification in the registry key by the ransomware to affect the behaviour of notifications ]
"HKLM\\SOFTWARE\\Microsoft\\WindowsNT\\CurrentVersion\\Notifications\\Data\\418A073AA3BC3475" \end{lstlisting}

\begin{lstlisting}[language=plantuml, caption=Modification in the registry key by the ransomware to replace the wallpaper ]
HKU\\S-1-5-21-4252838199-3690154089-1328794218-1003\\ControlPanel\\Desktop\\Wallpaper	"C:\\Users\\Support\\AppData\\Local\\Temp\\4506.tmp.bmp" 
\end{lstlisting}

\subsubsection{Network Connections}
Once the encryption got completed, the malware sample sent a request to https://iplogger.com. IPLogger is an IP address location tracking service. The attacker can create a tracker URL and when the malware sample connects to the URL, IPLogger tracks and logs the location of the infected device. Iplogger.com is currently hosted on Cloudflare's CDN network. In addition, there are several DNS queries, including requests to resolve pointer record (PTR) addresses and domains such as ‘c.pki.goog/r/gsr1.crl’ that may be associated with certificate verification activities or an attempt to avoid detection.

\begin{figure*}
\centering
\includegraphics[width=0.7\textwidth, height=7cm]{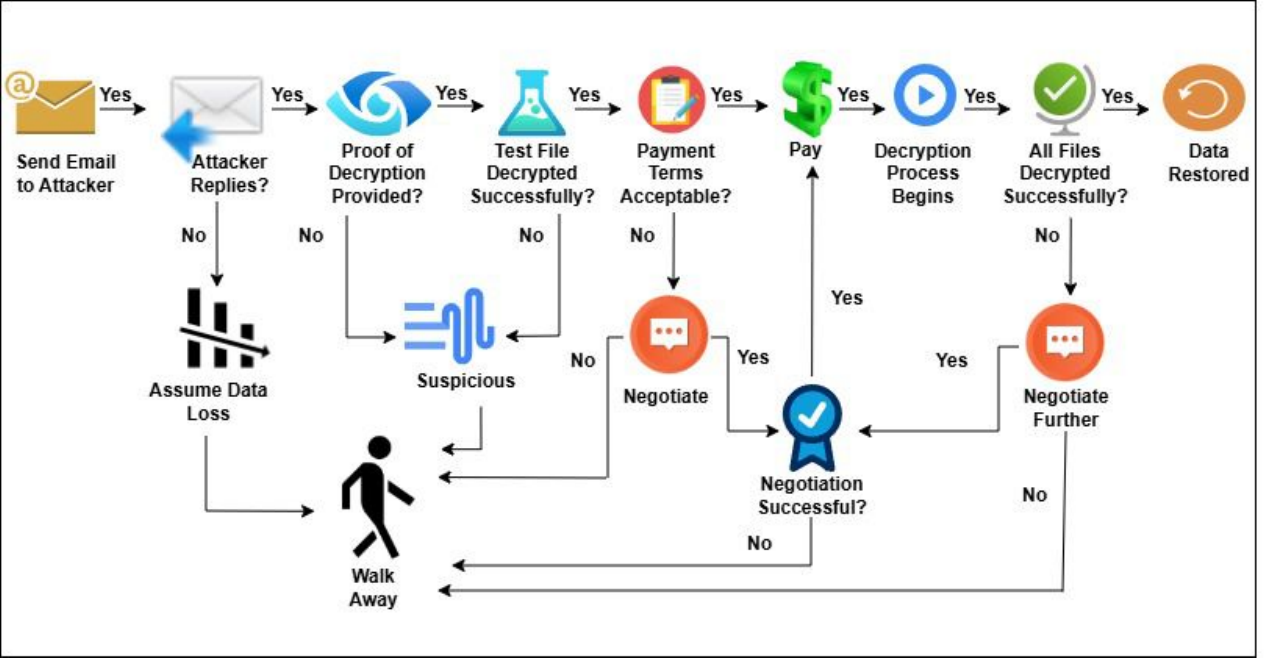}
  \caption{Plan to communicate with the attacker.}
  \label{fig:email_plan}
\end{figure*}

\subsection{Communication with the Attacker}
\label{subsec:comm_ransom}
The purpose of the honeypot deployment was to learn and gain information on threats related to water plants. Hence, we decided to proceed further and engage in one-to-one communication with the attacker. The attacker left a ransomware note, We communicated with the attacker using the contact information given in the ransomware note. The following assumptions could be made while communicating with the ransomware attacker, such as lowering the ransom cost, obtaining proof of decryption capability, identifying and understanding the scope of the attack, and verifying the legitimacy of the attacker (e.g., it is possible that it might be a scam).

We decided to follow up with the attacker to learn more about ransomware attacks and the intentions of the attacker. For this purpose, we first create a plan to communicate with the attacker in the form of decision tree as shown in Figure~\ref{fig:email_plan}. Following the plan, an initial email was sent to the attacker. This message was drafted to present the incident as an employee's oversight, with the employee appealing for file decryption to save his employment. To ascertain the attacker's decryption capability, two encrypted files were also attached. No mention of payment was made in this initial communication. Following the initial email, the attacker sent three responses. The first response outlined the attacker's demand in USD for file decryption. The attacker instructed us not to alter the encrypted files, including their extensions. The attacker provided his Bitcoin wallet address and stated that the decryption process would be shared upon payment. A decrypted version of the files we had sent was also provided.
A second email was sent by the attacker to confirm whether we had received his initial message. Later, attacker sent a third email to inquire about our continued interest in file decryption. Subsequent to these three responses from the attacker,  we drafted our second email. In this communication, we conveyed our continued interest in file decryption. However, the provided decrypted files were deemed insufficient to verify the attacker's comprehensive decryption capability; consequently, we forwarded additional files for decryption. We further expressed that the attacker's demand is huge and sought its reduction. Additionally, we indicated our unfamiliarity with the Bitcoin payment process and requested alternative payment method. After our second email, the attacker sent two more responses. In these responses, the attacker provided a web link to access the decrypted files. The attacker also reduced his demand to 3750 USD. However, the attacker insisted on using Bitcoin as the sole payment method. Subsequently, the attacker sent another email to follow up, stating that they could offer a further discount until the end of the weekend. Following this interaction, we stopped communication with the attacker, having obtained adequate information about the ransomware attacker and the attack. Also, we did not want to follow the link given by the attacker for safety purposes.

\section{Discussion}
\label{sec:discussion}

\subsection{Data backup and recovery}
The implementation of an efficient data backup and recovery plan is crucial for critical systems. For example, in this honeypot scenario,  a dedicated workstation was established for data backup. This workstation was intentionally isolated from the main honeypot network by blocking the coming traffic while allowing outgoing traffic. As shown in Figure~\ref{fig:honeywin_dt},  workstation (U2) was setup as a backup server to save the incoming network traffic captured in U0, the outgoing network traffic captured via the SPAN port with U1, successful logins and failed login attempts, and Elastic search indices of Sysmon events.  U2 connects to U0 and U1 on the Backup Network and does not have Internet connection. Also, only U2 has access to U0 and U1 but not vice versa.

\subsection{Pros and Cons of Communicating with the Ransomware Attacker}
Though one may not recommend  communication with a ransomware attacker due to perceived potential risks, we did so as a research study. After the ransomware attack, we decided to proceed further and engaged in one-to-one communication with the attacker. The assumptions we discussed in Section~\ref{subsec:comm_ransom} were found to be true. The attacker did lower the  ransom payment. We also noticed a sense of urgency in the attacker during the communication. The communication also highlighted the potential risks of communicating with the attacker; for example, for decryption purposes, the attacker sent a link to access the files instead of sending them directly as an attachment. Opening a link naturally carries potential risks. Moreover, the attacker was relying on Bitcoin payments rather than traditional banking channels. The attacker was extremely cautious regarding the potential damage to our encrypted files, as can be seen in the ransomware note. The attacker also indicated the intention to reduce the ransom payment further, which ultimately reveals the now well known  purpose and intentions in launching such ransomware attacks.

\subsection{Security improvements}
The proposed honeypot setup was designed to attract attackers, however it underscores the critical importance of implementing stronger security controls. These controls should include multi-factor authentication to prevent unauthorized access in case of compromised credentials, and guidelines for employees on the secure use of critical systems accessed via remote desktop or other similar platforms to mitigate the human-centric vulnerabilities.

\section{Conclusion}
\label{sec:conclusion}
A water treatment plant-based cyber twin was deployed as a honeypot to assess threats to water plants. Remote Desktop Protocol (RDP) (i.e., Port 3389) and Secure Shell (SSH) (i.e., Port 22) were intentionally left open to create vulnerabilities. These vulnerabilities were further exacerbated by the use of a weak password. This setup was successfully attacked by a ransomware attacker. However, due to the careful implementation of the entire setup, the attacker was unable to inflict significant damage. All data was securely backed up to a separate workstation. We also engaged in one-to-one communication with the ransomware attacker to gain further insights into threats. All communication conducted with the ransomware attacker is also reported in this study to provide a comprehensive understanding of threats associated with critical infrastructures.

%

%

%

\section*{Acknowledgement}
This research is supported in part by the National Research Foundation, Singapore, under its National Satellite of Excellence Programme “Design Science and Technology for Secure Critical Infrastructure: Phase II” (Award No: NRF-NCR25-NSOE05-0001). Any opinions, findings and conclusions or recommendations expressed in this material are those of the author(s) and do not reflect the views of National Research Foundation, Singapore. The authors acknowledge the assistance provided by Shreyas Ravishankar, Sebastian Tay Yong Xun, and Tay Jovan in developing and deploying the honeypot. We thank  Nagarajan SIVANADIPATHAM for providing Figure\,\ref{fig:hmi}.


\bibliographystyle{unsrt}
\balance
\bibliography{references.bib}
\end{document}